# Average Doppler Shift of Gamma-ray Spectra of Positron Annihilation Process in Molecules

Fang Yuan, Xiaoguang Ma*, Yu Wu, and Chuanlu Yang

*School of Physics and Optoelectronic Engineering, Ludong University, Yantai, Shandong, 264025, People's Republic of China*

This paper studies the gamma-ray spectra of positron annihilation processes in a series of molecules. The results show that the average valence electron energy of the molecules has a linear correlation with the full width at half maximum (FWHM) of the gamma-ray spectra. In addition, we defined a new physical quantity Average Doppler Shift (ADS), which can be used as the eigenvalue to describe the characteristics of the gamma-ray spectra. Since ADS contains all the information about the gamma-ray spectra, it can more accurately represent the characteristics of the gamma-ray spectra. For a series of molecules, this paper compares the ADS and FWHM of their gamma-ray spectra and the average valence electron energy. The results show that ADS has a linear correlation with the average valence electron energy and the FWHM. Further, this proves that the annihilation mainly occurs on valence electrons, and it also illustrates that the ADS has certain applicability. It is expected that this will provide us with a deeper understanding of the positron annihilation process.

## 1. Introduction

Positron is the most basic antiparticle, it will annihilate with electrons and release gamma-rays. In recent years, the research on the gamma-ray spectra in the process of positron annihilation has never stopped.[1-6] The potential exerted by the atomic nucleus on a positron is repulsive so that the amplitude of the positron wave function is small near the inner-shell electrons.[7] The positrons predominantly annihilate with the valence electrons in atoms or molecules. Most of the experimental measurement results are also consistent with the theoretical gamma-ray spectra of valence electrons.[7-10] Obviously, valence electrons dominate the positron annihilation process. However, almost no research can give a physical quantity that depends on the valence electron to quantitatively illustrate that the annihilation electron is the valence electron in the positron annihilation process.

The full width at half maximum (FWHM), a physical parameter, has long been used to describe gamma-ray spectra. [11-15] The width FWHM is only an experimental spectra analysis parameter. The width FWHM of the gamma-ray spectra is obtained by a Two-Gaussian function fitting. This fitting function has no physical significance of which we are aware, and this functional form serves the purpose of representing the experimentally measured line shapes analytically with reasonable accuracy. [7] FWHM has never had a clear physical meaning, which makes it difficult for us to fully understand the mechanism of the positron annihilation process.

In our previous research on the gamma-ray spectra of positron annihilation in the atom, we found that there is a linear correlation between the atomic Rahm's electronegativity [16] and the FWHM of the gamma-ray spectra in the atoms. Since Rahm regards electronegativity as the average energy of the valence electrons in the ground state, it is obviously closely related to the valence electrons. The linear relationship between the two partly explains the dominance of valence electrons in the annihilation process. In addition, we propose a new physical quantity Average Doppler Shift (ADS), which is derived from the integral of the entire gamma-ray spectra. The ADS can describe all the characteristics of the gamma-ray spectra accurately and comprehensively. There is a good linear relationship between ADS, average energy of valence electrons and FWHM, which also proves that ADS has certain applicability.

However, in previous studies on a range of atoms, both FWHM and ADS, the data were derived from theoretical calculations. Because there is no experimental data on atoms other than noble gas atoms, our conclusions cannot be experimentally verified. Recently, the Doppler-broadened gamma-ray spectra for low-energy positron annihilation in many gas-phase molecules have been measured extensively with milestone achievements, which provides strong support for our research. [17-20] Therefore, in this paper, we study the positron annihilation gamma ray spectra of a series of molecules, and used the experimental values obtained by the Gaussian experiment fitting of 59 molecules provided by the C.M.Surko's group.[7] The theoretical calculation of this study only includes the valence electrons of the molecule. Through the comparative study of experimental values and theoretical values, some explanations and opinions are put forward.

Electronegativity is usually used to express the ability of an atom to attract electrons in a compound. It is a scale for atoms, and there are many calculation methods. In the definition of Rahm's electronegativity, electronegativity is regarded as the average energy of ground state valence electrons. Since there is no known electronegativity data for molecules, we calculated the average energy of the valence electrons in a series of molecules and regarded it as a scale equivalent to electronegativity.

Reapplying our previous research work on atoms to molecules can prove experimentally that our conclusion is correct, thus further proving that positrophilic electrons in the process of positron annihilation are valence electrons, and further illustrating the general applicability of ADS.

## 2. Theoretical introduction

The gamma-ray spectrum in the process of positron-electron annihilation is usually measured by the momentum distribution of the positron and electron pairs. [10]

$$A_{i\vec{k}}(\vec{P}) = \int \psi_i(\vec{r})\phi_{\vec{k}}(\vec{r})\, e^{-i\vec{P}\cdot\vec{r}}\, d\vec{r} \qquad (1)$$

where $\psi_i(\vec{r})$ and $\phi_{\vec{k}}(\vec{r})$ are the wavefunctions for electrons and positrons respectively.

In order to study the relationship between momentum spectra and gamma-ray spectra, the radial distribution function of momentum space is defined. [15]

$$D(p) = \int_0^\pi d\theta \int_0^{2\pi} d\phi\, P^2 \sin\theta |A_i(\vec{P})|^2 \qquad (2)$$

where P, $\theta, \phi$ are spherical coordinates respectively. Hence, the theoretical spherically averaged momentum distribution is [15]

$$\sigma_i(P) = \frac{D_i(P)}{4\pi P^2} \qquad (3)$$

Equation (3) represents the average probability of electron-positron pairs appearing on the surface with momentum | P |. [15]

The gamma-ray spectra in the annihilation process is then Doppler shifted in energy due to the longitudinal momentum component of positron-electron pair. Therefore, to obtain the total probability density when the momentum is P = 2ϵ / c, the plane perpendicular to P must be integrated. Then the gamma-ray spectra for positron-electron pair are [15]

$$\Omega_i(\epsilon) = \frac{1}{c}\int_{2\epsilon/c}^\infty \sigma_i(P) P\, dP \qquad (4)$$

The central Doppler shift ($mc^2 = 511\text{keV}$) is given by $\epsilon$. For low-energy positrons, its energy spectrum is determined by the bound electrons in the molecule. The Doppler shift $\epsilon$ is directly related to the binding energy $\epsilon_n$ of the annihilation bound electrons, i.e., $\epsilon \propto \sqrt{\epsilon_n}$ .[11] In other words, the Doppler shift is determined by the molecular bond energy. If we integrate the gamma-ray spectrum, the annihilation rate is [12]

$$Z_{eff} = \int \Omega(\epsilon) d\epsilon \qquad (5)$$

The observation spectrum obtained by the Two-Gaussian fitting is [7]

$$\Omega(\epsilon) = \exp\left[-\left(\frac{\epsilon - \epsilon_0}{a\Delta\epsilon_1}\right)^2\right] + D\exp\left[-\left(\frac{\epsilon - \epsilon_0}{a\Delta\epsilon_2}\right)^2\right] \qquad (6)$$

where $a = \frac{1}{(4\ln 2)^{1/2}}$ , $\epsilon_0 = 0$ , $\Delta\epsilon_1$, $\Delta\epsilon_2$, D were all data provided by C.M.Surko's group.

Then the average bonding energy of annihilated electrons (actually the Average Doppler Shift) will be

$$\bar{\epsilon} = \frac{\int \Omega(\epsilon)\epsilon d\epsilon}{\int \Omega(\epsilon) d\epsilon} \qquad (7)$$

The gamma-ray spectra carries information about the distribution of electron momentum in the orbit of the bound state.[23] In this study, $\bar{\epsilon}$ is called the Average Doppler Shift (ADS). The Doppler shift of the gamma-ray spectra is proportional to the square root of the absolute value of the bound energy of the annihilated bound electrons.[10] Therefore, the ADS of the gamma-ray spectra represents the average bonding energy of positron pairs.

Rahm's new electronegativity scale is [16]

$$\bar{\chi} = \sum_{i=1}^{n} \frac{n_i \epsilon_i}{n} \qquad (8)$$

where $\epsilon_i$ is the energy of the $i$th level, $n_i$ is the occupation the $i$th level, and $n$ is the total number of electrons. Electronegativity is regarded as the ground-state average valence electron binding energy, and ADS is regarded as the average binding energy of annihilation electrons. By comparing the ADS and Rahm's electronegativity values of gamma-ray spectra, the dominant role of valence electrons in positron annihilation can be proved quantitatively.

## 3. Results and Discussion

In this section, the average valence electron energy, the experimental and theoretical values of FWHM and ADS of the positron annihilation gamma-ray spectra in a series of molecules are calculated (see Appendix Table 1 for specific data), and these data are analyzed and fitted. In this study, we regard the average valence electron energy of a molecule as a scale equivalent to the electronegativity of the atoms. Atomic electronegativity is a scale indicating the tendency of atoms in a molecule to attract electrons. [24] In this study, its exact definition is considered to be its ability to attract positrons.

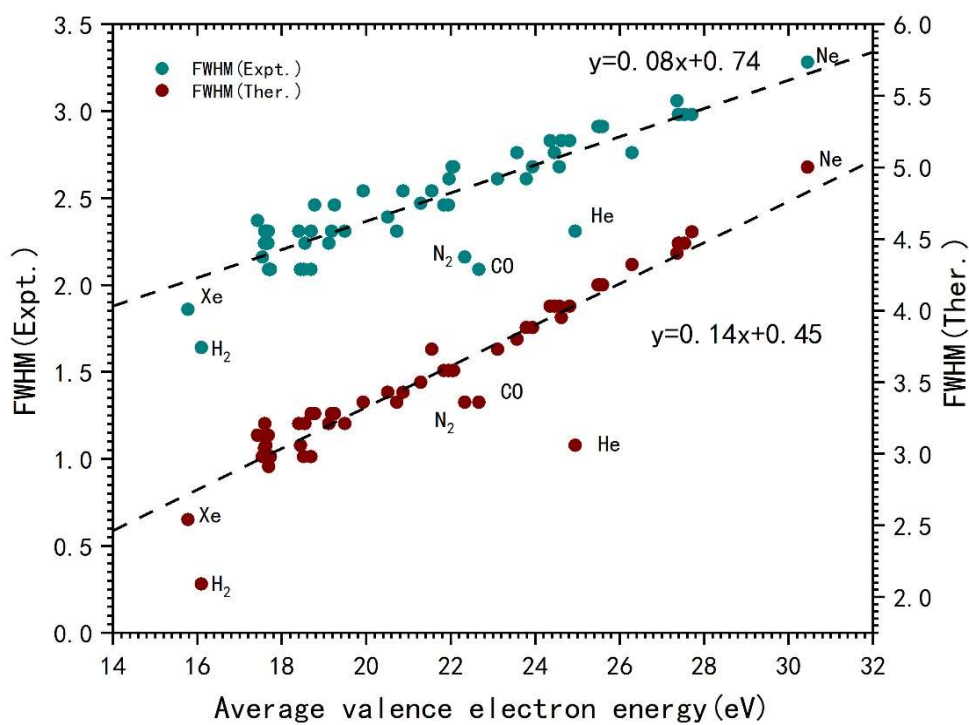

Fig. 1. The fitting diagram of average valence electron energy with theoretical values and experimental values of the FWHM, respectively

We calculated the average valence electron energy of 59 molecules based on Rahm's definition of electronegativity. Since there are no experimental values for comparison in atomic research, it is impossible to prove our conclusions through experiments. Therefore, we fit the average valence electron energy of 59 molecules with the FWHM experimental values and theoretical values of their gamma-ray spectra, as shown in Fig.1. The experimental values are obtained by the C.M. Surko group using Two-Gaussian fitting experiments. Their measurement data is accurate to study the shape of the spectra line, not just the width. It is obviously that there is a strong linear correlation between the average valence electrons energy and the theoretical value of the FWHM, which is consistent with our results in atoms. In addition, it is also observed that the average valence electrons energy has a good linear relationship with the experimental value of FWHM, which proves our conclusion that FWHM can be used to represent the ability of valence electrons in molecules to attract positrons. The ratio of the average valence electron energy to the experimental value of FWHM is 12.5, and that to the theoretical value of FWHM is 7.14. It was observed that $N_2$, CO and He were free outside the fitting line, which was very consistent with the experimental results of C.M. Surko's group. In their study, $N_2$, CO and Helium showed strong non-Gaussian features.[7]

In previous studies, FWHM has been used to express the characteristics of gamma-ray spectra. It is well known that FWHM is just a special spot on the gamma-ray spectra, and it has no known physical significance. However, gamma-ray spectra usually have more than one peak and shoulder, so FWHM cannot fully represent all characteristics of gamma-ray spectra. According to the definition of Rahm's electronegativity, we propose a new physical quantity Average Doppler Shift, which can also be used to express the characteristics of the gamma-ray spectra. ADS comes from integrating the whole gamma-ray spectra, that is, it contains all the information of the gamma-ray spectra.

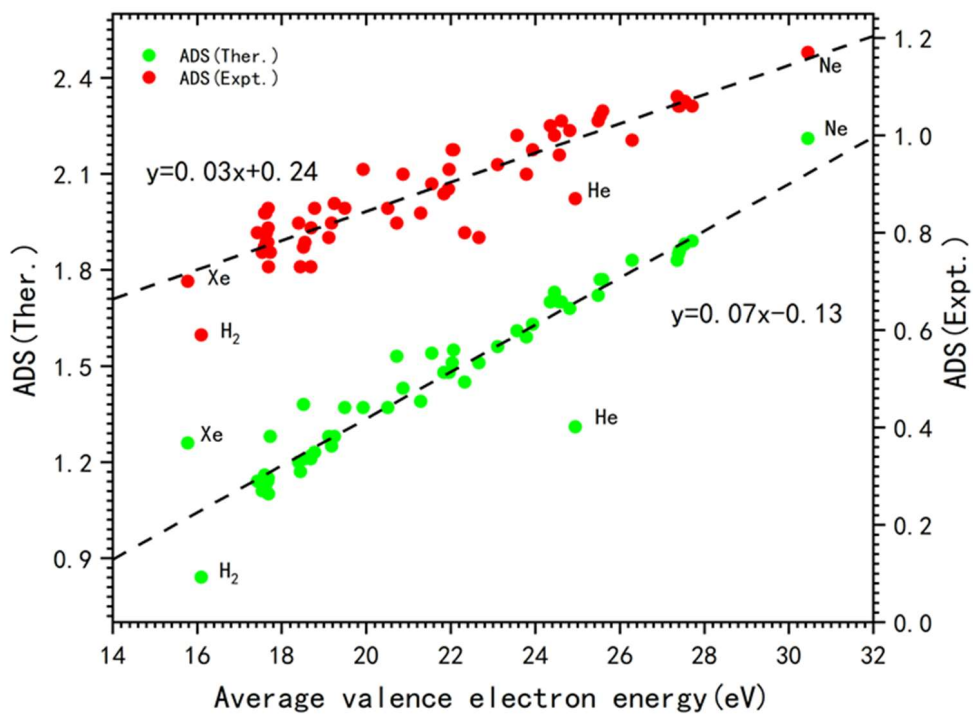

Fig. 2. The fitting diagram of average valence electron energy with theoretical values and experimental value of the ADS, respectively

We calculated the theoretical and experimental values of ADS for 59 molecules. In Fig.2, we linearly fit the average valence electrons energy and the theoretical and experimental values of ADS, and it is found that the average valence electrons energy has an obvious linear relationship with the theoretical value of ADS. This is consistent with the conclusion we reached when we studied atoms. When calculating the theoretical value of ADS, we use the gamma-ray spectra data of the valence electrons orbital. Therefore, it is easy to infer the linear relationship between the theoretical value of ADS and the average valence electrons energy. However, in this study, it is found that the experimental value of ADS is also linearly correlated with the average valence electron energy. The ADS is regarded as the average binding energy of the annihilated electrons, so it can be proved quantitatively that positrophilic electrons are valence electrons.

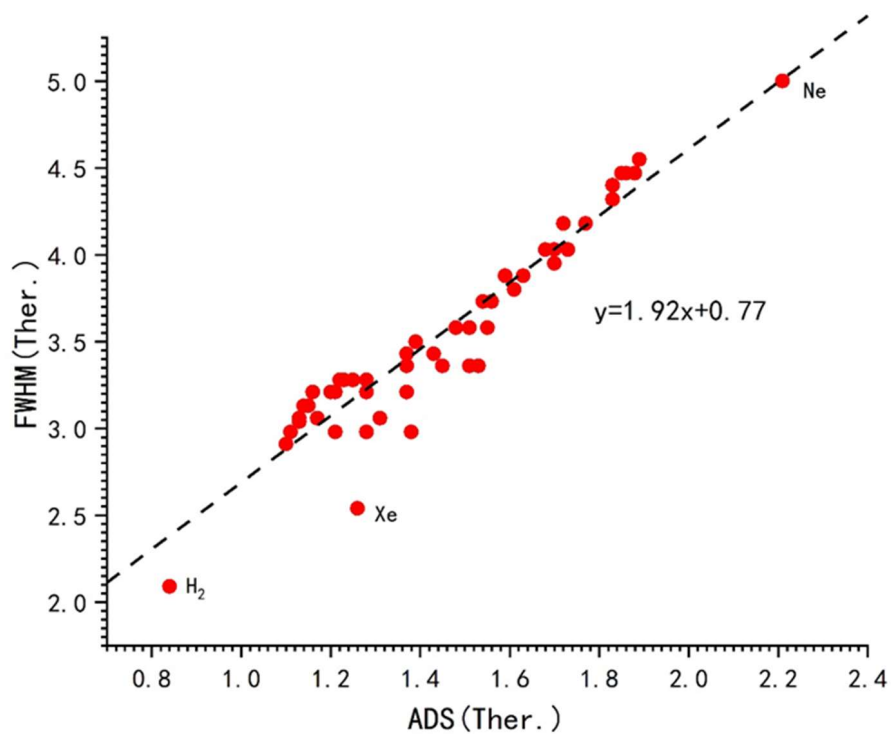

Fig. 3. Linear fitting diagram of ADS theoretical values and FWHM theoretical values

    We linearly fit the theoretical value of ADS and FWHM. In Fig.3, it can be seen that the theoretical value of ADS and the theoretical value of FWHM have a strong linear correlation. In theory, ADS can be used as a reference value for describing gamma ray spectra. To further prove our results, we fit the ADS experimental value and FWHM experimental value. Both sets of data are obtained by the Two-Gaussian fitting experiments. Such as Fig. 4, obviously, the experimental values fit very well, but there are still molecules free outside the line. Combined with Fig.3, we can infer that FWHM loses some information of positron annihilation gamma spectra. The FWHM is only a point of the gamma ray spectra, the error of FWHM increases. The ADS includes all points of gamma ray spectra, and all the information of gamma-ray spectra is on ADS. Therefore, ADS can more accurately represent the characteristics of the gamma-ray spectra.

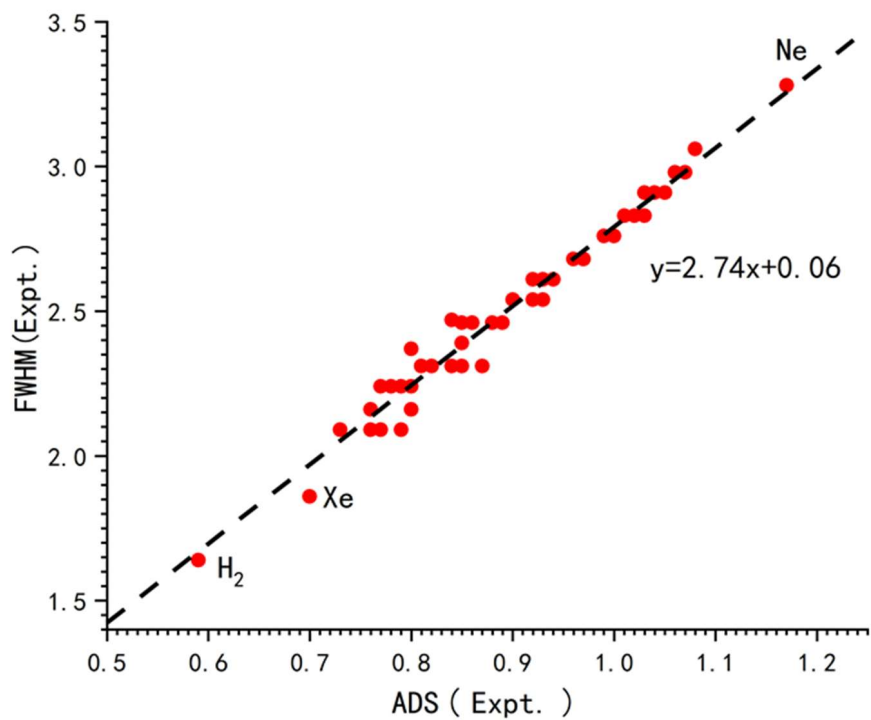

Fig. 4. Linear fitting diagram of ADS experimental values and FWHM experimental values

The inert gases, alkane molecules, halogenated and partial hydrocarbons, and perfluorinated hydrocarbons in 59 molecules were separately listed for further analysis. In Fig. 5, the molecular in I area are the inert gas molecules. Their ADS experimental value and theoretical value and FWHM experimental value and theoretical value change consistent. For noble gases, in addition to helium molecules, the ADS and FWHM of several other molecules decrease with the increase of the number of electron layers, and helium is relatively special. We think this is caused by its relatively special electron orbital structure. Helium has only one layer of electrons, so it is difficult to distinguish core electrons from valence electrons. The molecules in II area are alkane molecules. Obviously, their ADS experimental values and FWHM experimental values have the same changing trend. The theoretical value of ADS and FWHM have the same change trend. But the difference between ADS and FWHM between each molecule is not big. The FWHM and ADS don't increase with the increase of C - C bond and C - H bond. It shows that the alkane positron annihilation of the gamma-ray spectra is similar.

The molecules in III area are halogenated hydrocarbon molecules, They are $CF_4$, $CCl_4$ and $CBr_4$. Their ADS and FWHM are declining. In our research on atoms, we know that the electronegativity of F, Cl, and Br gradually decreases. This gives us reason to believe that electronegativity is related to positron annihilation. In IV, V, VI, VII, there are a series of fluorinated hydrocarbon molecules. The number of C atoms in each group of fluorinated hydrocarbon molecules is the same, and the number of F atoms replacing H atoms increases sequentially. Several groups of data show the same characteristics that with the increase of F atoms, ADS and FWHM also gradually increase. This is consistent with the findings of Tang and Iwata et al. on these molecules, the annihilation site is on the F atom.[22] Positrons annihilate halogen atoms.[7]

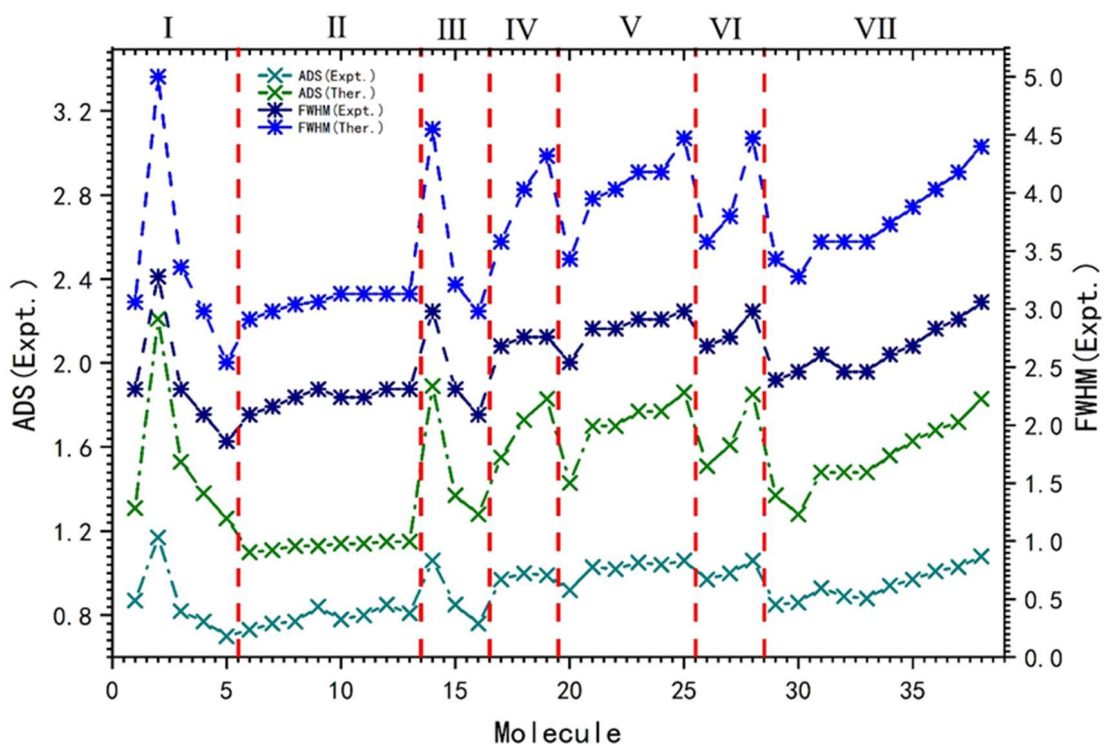

Fig.5 Comparison diagram of ADS and FWHM experiments and theoretical values of inert gases, halogenated hydrocarbons and partially and fully fluorinated hydrocarbons

## 4. Conclusions

We have studied the experimental and theoretical values of ADS and FWHM and the average energy of valence electrons in 59 molecules. There is a good linear relationship between the average valence electron energy and the experimental and theoretical values of ADS and FWHM. It shows that in the process of positron annihilation, valence electrons in the molecule is playing the main role. On the other hand, it shows that ADS can be used as a reference value to describe the characteristics of the gamma-ray spectra. The ADS contains all the information about the gamma-ray spectra in contrast to the FWHM, which is only a feature point of the gamma spectra. The ADS shows advantages in describing gamma-ray spectra. The ADS has a definite physical meaning. The ADS is the Average Doppler shift, which represents the bonding ability of positron pairs. It is helpful to study the mechanism of the positron annihilation process.

**Acknowledgment**

The authors acknowledge the support of the National Natural Science Foundation of China under grant no. 11674145 and Taishan Scholars Program of Shandong Province (ts2015110055).

*E-mail: hsiaoguangma@ldu.edu.cn

**APPENDIX**

The theoretical calculations are mainly implemented by using the Gaussian09 computational chemistry package. In the ab initio Hartree Fock calculations, the TZVP basis set13) is used, i.e., the calculation scheme is the HF/ TZVP scheme. The molecular wavefunctions are then obtained using the HF/TZVP scheme and optimised structural parameters.

TABLE I. ADS theoretical and experimental values, FWHM theoretical and experimental values, and average valence electron energy

| Molecule | number | ADS (Expt.) | ADS (Ther.) | FWHM (Expt.) | FWHM (Ther.) | Average valence electron energy |
|---|---|---|---|---|---|---|
| He | 1 | 0.87 | 1.31 | 2.31 | 3.06 | 24.95 |
| Ne | 2 | 1.17 | 2.21 | 3.28 | 5.00 | 30.45 |
| Ar | 3 | 0.82 | 1.53 | 2.31 | 3.36 | 20.72 |
| Kr | 4 | 0.77 | 1.38 | 2.09 | 2.98 | 18.52 |
| Xe | 5 | 0.7 | 1.26 | 1.86 | 2.54 | 15.78 |
| $H_2$ | 6 | 0.59 | 0.84 | 1.64 | 2.09 | 16.10 |
| $N_2$ | 7 | 0.80 | 1.45 | 2.16 | 3.36 | 22.33 |
| $O_2$ | 8 | 0.96 | 1.7 | 2.68 | 4.03 | 24.57 |
| CO | 9 | 0.79 | 1.51 | 2.09 | 3.36 | 22.67 |
| $CO_2$ | 10 | 0.92 | 1.59 | 2.61 | 3.88 | 23.79 |
| $H_2O$ | 11 | 0.90 | 1.54 | 2.54 | 3.73 | 21.55 |
| $SF_6$ | 12 | 1.07 | 1.88 | 2.98 | 4.47 | 27.54 |
| $NH_3$ | 13 | 0.79 | 1.28 | 2.24 | 3.21 | 19.12 |
| $CH_4$ | 14 | 0.73 | 1.1 | 2.09 | 2.91 | 17.69 |
| $C_2H_6$ | 15 | 0.76 | 1.11 | 2.16 | 2.98 | 17.54 |
| $C_3H_8$ | 16 | 0.77 | 1.13 | 2.24 | 3.04 | 17.59 |
| $C_4H_{10}$ | 17 | 0.84 | 1.13 | 2.31 | 3.06 | 17.62 |
| $C_5H_{12}$ | 18 | 0.78 | 1.14 | 2.24 | 3.13 | 17.63 |
| $C_6H_{14}$ | 19 | 0.80 | 1.14 | 2.24 | 3.13 | 17.64 |
| $C_9H_{20}$ | 20 | 0.85 | 1.15 | 2.31 | 3.13 | 17.68 |
| $C_{12}H_{26}$ | 21 | 0.81 | 1.15 | 2.31 | 3.13 | 17.68 |
| $C_6H_{12}$ | 22 | 0.84 | 1.16 | 2.31 | 3.21 | 17.60 |
| $CH_3C(CH_3)HC_2H_5$ | 23 | 0.78 | 1.14 | 2.24 | 3.13 | 17.66 |
| $C(CH_3)_4$ | 24 | 0.78 | 1.14 | 2.24 | 3.13 | 17.67 |
| $C_2H_4$ | 25 | 0.73 | 1.17 | 2.09 | 3.06 | 18.44 |
| $C_2H_2$ | 26 | 0.73 | 1.21 | 2.09 | 2.98 | 18.69 |
| $C_6H_6$ | 27 | 0.78 | 1.21 | 2.24 | 3.21 | 18.55 |
| $C_{10}H_8$ | 28 | 0.81 | 1.22 | 2.31 | 3.28 | 18.70 |
| $C_{14}H_{10}$ | 29 | 0.85 | 1.23 | 2.46 | 3.28 | 18.78 |
| $C_6H_5CH_3$ | 30 | 0.82 | 1.20 | 2.31 | 3.21 | 18.40 |
| $CF_4$ | 31 | 1.06 | 1.89 | 2.98 | 4.55 | 27.71 |
| $CCl_4$ | 32 | 0.85 | 1.37 | 2.31 | 3.21 | 19.49 |
| $CBr_4$ | 33 | 0.76 | 1.28 | 2.09 | 2.98 | 17.73 |
| $CH_3F$ | 34 | 0.97 | 1.55 | 2.68 | 3.58 | 22.07 |
| $CH_2F_2$ | 35 | 1.00 | 1.73 | 2.76 | 4.03 | 24.46 |
| $CHF_3$ | 36 | 0.99 | 1.83 | 2.76 | 4.32 | 26.29 |
| $C_2H_5F$ | 37 | 0.92 | 1.43 | 2.54 | 3.43 | 20.87 |
| $CF_3CH_3$ | 38 | 1.03 | 1.7 | 2.83 | 3.95 | 24.62 |
| $CHF_2CH_2F$ | 39 | 1.02 | 1.7 | 2.83 | 4.03 | 24.36 |
| $CF_3CH_2F$ | 40 | 1.05 | 1.77 | 2.91 | 4.18 | 25.60 |
| $CHF_2CHF_2$ | 41 | 1.04 | 1.77 | 2.91 | 4.18 | 25.54 |
| $C_2F_6$ | 42 | 1.06 | 1.86 | 2.98 | 4.47 | 27.42 |
| $CH_3CF_2CH_3$ | 43 | 0.97 | 1.51 | 2.68 | 3.58 | 22.04 |
| $CF_3C_2H_5$ | 44 | 1.00 | 1.61 | 2.76 | 3.80 | 23.57 |
| $C_3F_8$ | 45 | 1.06 | 1.85 | 2.98 | 4.47 | 27.39 |
| $CH_2FC_5H_{11}$ | 46 | 0.86 | 1.28 | 2.46 | 3.28 | 19.25 |
| $C_6H_5F$ | 47 | 0.85 | 1.37 | 2.39 | 3.43 | 20.51 |
| $C_6H_4F_2$-2 | 48 | 0.93 | 1.48 | 2.61 | 3..58 | 21.96 |
| $C_6H_4F_2$-3 | 49 | 0.89 | 1.48 | 2.46 | 3.58 | 21.95 |
| $C_6H_4F_2$-4 | 50 | 0.88 | 1.48 | 2.46 | 3.58 | 21.84 |
| $C_6H_3F_3$ | 51 | 0.94 | 1.56 | 2.61 | 3.73 | 23.11 |
| $C_6H_2F_4$ | 52 | 0.97 | 1.63 | 2.68 | 3.88 | 23.94 |
| $C_6HF_5$ | 53 | 1.01 | 1.68 | 2.83 | 4.03 | 24.82 |
| $C_6F_6$ | 54 | 1.03 | 1.72 | 2.91 | 4.18 | 25.49 |
| $C_6F_{14}$ | 55 | 1.08 | 1.83 | 3.06 | 4.40 | 27.36 |
| $CH_3OH$ | 56 | 0.93 | 1.37 | 2.54 | 3.36 | 19.93 |
| $Si(C_2H_5)_4$ | 57 | 0.80 | 1.14 | 2.37 | 3.13 | 17.43 |
| $C_6H_5NO_2$ | 58 | 0.84 | 1.39 | 2.47 | 3.50 | 21.29 |
| $C_5H_5N$ | 59 | 0.82 | 1.25 | 2.31 | 3.28 | 19.18 |